\documentclass[
	reprint,
    aps,prl,superscriptaddress,
    floatfix
   ]{revtex4-2}

\newcommand{\etal}{{\em et al.}}

\newcommand{\LigoMIT}{LIGO, Massachusetts Institute of Technology, Cambridge, MA 02139, USA}
\newcommand{\LigoCIT}{LIGO, California Institute of Technology, Pasadena, CA 91125, USA}

\usepackage{xcolor}
\usepackage{graphicx}
\usepackage{amsfonts}
\usepackage{amsmath}
\usepackage{bm}
\usepackage{hyperref}
\usepackage{cleveref}
\usepackage{mathtools}
\usepackage{physics}

\hypersetup{
colorlinks=true,
linkcolor=blue,
citecolor=blue,
urlcolor=blue
}
\usepackage[
        retain-explicit-plus=true,
        retain-unity-mantissa=false,
        range-phrase=-,
        range-units=single
    ]{siunitx}
\graphicspath{{figures/}}

\newcommand{\ee}{\mathrm{e}}
\renewcommand{\t}[1]{\mathrm{#1}}

\begin{document}

\title{Unification of thermal and quantum noise in gravitational-wave detectors}

\author{Chris Whittle}
\email{chris.whittle@ligo.org}
\affiliation{\LigoMIT}
\author{Lee McCuller}
\affiliation{\LigoCIT}
\author{Vivishek Sudhir}
\email{vivishek@mit.edu}
\affiliation{\LigoMIT}
\affiliation{Department of Mechanical Engineering, Massachusetts Institute of Technology, Cambridge, MA 02139}
\author{Matthew Evans}
\affiliation{\LigoMIT}

\date{\today}

\begin{abstract}
    Contemporary gravitational-wave detectors are fundamentally limited by thermal noise---due to dissipation in the mechanical elements of the test mass---and quantum noise---from the vacuum fluctuations of the optical field used to probe the test mass position.
    Two other fundamental noises can in principle also limit sensitivity: test-mass quantization noise due to the zero-point fluctuation of its mechanical modes, and thermal excitation of the optical field.
    We use the quantum fluctuation-dissipation theorem to unify all four noises.
    This unified picture shows precisely when test-mass quantization noise and optical thermal noise can be ignored.
\end{abstract}

\maketitle

\emph{Introduction.}---Fundamental constraints on the sensitivity of gravitational-wave (GW) detectors arise from classical and quantum fluctuations.
At present, each of these noises is modeled using different techniques.
Thermal noise---due to the Brownian motion of the mechanical test mass, its suspension, and the mirror coating on the test mass---is derived from the (classical) fluctuation-dissipation theorem~\cite{Saulson90,Levin98}.
Quantum noise---due to the vacuum fluctuations in the phase and amplitude of the optical field used to measure the test mass position---is derived from quantum electrodynamics in the so-called ``two-photon formalism''~\cite{Caves81,Caves85,Buonanno01,Kimble01,Corb05}.
The sum of these noises limit the performance of today's GW detectors: quantum fluctuations in the amplitude of the optical field drive the motion of the test mass in the $\SI{\sim 20}{}$-$\SI{50}{\hertz}$ range~\cite{Yu20,Acernese20}, Brownian motion of the mirror coatings dominate in the $\SI{\sim 50}{}$-$\SI{200}{\hertz}$~\cite{Buik20}, and quantum fluctuations in the phase of the optical field sets the sensitivity above $\SI{200}{\hertz}$~\cite{Tse19,Acernese19}.

In principle there exist noises that are exactly complementary, i.e. quantum noise of the mechanical degrees of freedom and thermal noise of the optical field.
The former is a consequence of quantizing the mechanical motion of the interferometer test masses and the zero-point fluctuations that manifest as a result.
In fact, Braginsky \etal{} studied the role of test-mass quantization noise~\cite{Braginsky03}, concluding that \emph{``test-mass quantization is irrelevant [\ldots] if one filters the output data appropriately''}.
On the other hand, thermal fluctuations of the optical field---for example due to blackbody radiation---can contribute excess noise.

We show that the four fundamental noises described above---thermal and quantum noises of the mechanical and optical degrees of freedom---can all be treated uniformly using the quantum extension of the fluctuation-dissipation theorem~\cite{CallWelt51,Kubo66,Lax60,CouRey92,Clerk10,Miao19}.
This perspective enables a simple treatment of the test-mass quantum noise that is independent of the detector topology, instead depending only on the relative thermal and optical quantum energy scales.
In doing so, we extend the analysis of Ref.~\cite{Braginsky03} to incorporate mechanical losses and to allow for differing GW detector optical topologies or arbitrarily complex test mass suspensions.
We find that test-mass quantization noise is negligible in principle---i.e. \emph{independent} of any ``filtering''---so long as the mechanical degrees of freedom of the test mass resonate at acoustic frequencies ($\Omega_m$) and the detector is operated at a temperature
\begin{equation*}
    T > \hbar \Omega_m/k_B \approx (\SI{5e-11}{\kelvin})\cdot \left(\frac{\Omega_m}{2\pi \cdot\SI{1}{\hertz}}\right).
\end{equation*}
Likewise, optical thermal noise at the carrier frequency $\omega_o$ is negligible compared to its quantum noise as long as GW detectors are operated at a temperature
\begin{equation*}
    T < h \omega_o/k_B \approx (\SI{14e3}{\kelvin})\cdot \left(\frac{\omega_o}{2\pi\cdot\SI{300}{\tera\hertz}}\right).
\end{equation*}

\begin{figure*}[t!]
    \includegraphics{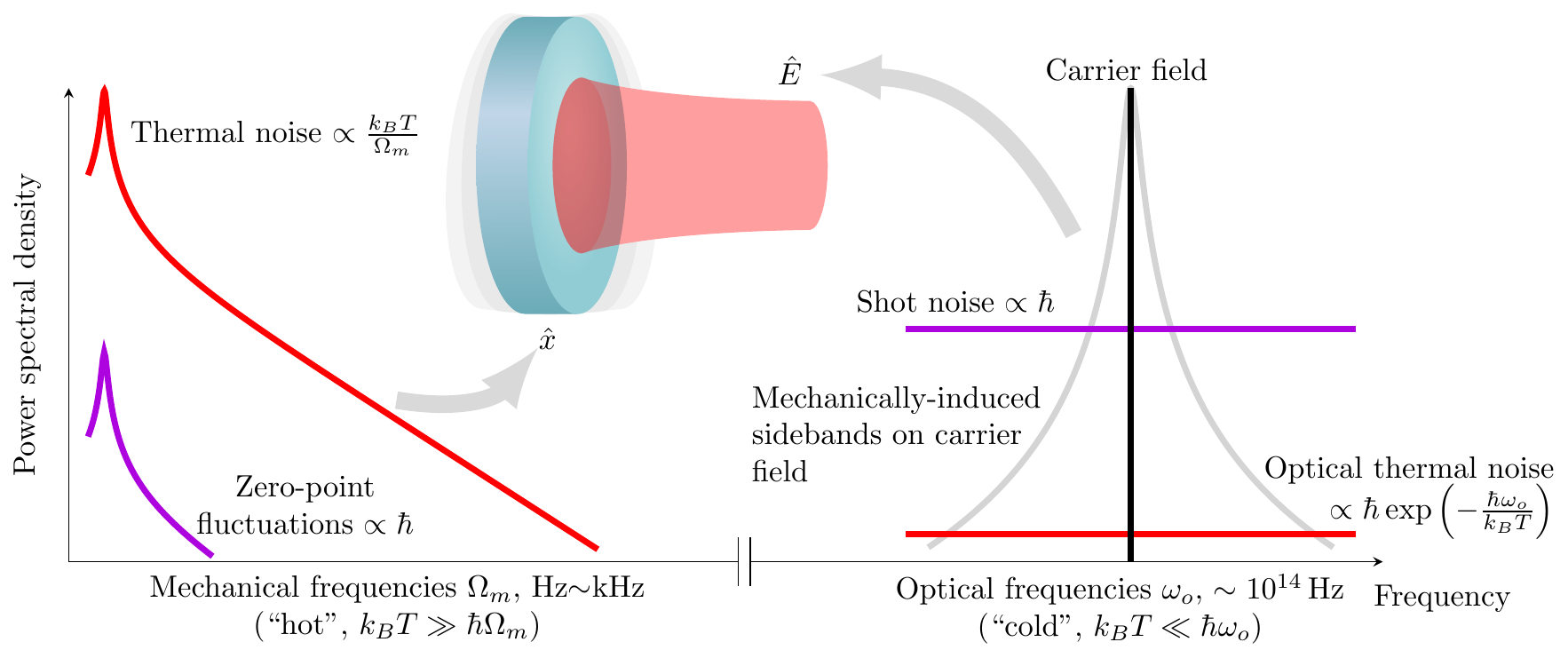}
    \caption{\label{fig:spectrum}
        A qualitative depiction of noise terms arising from the general fluctuation-dissipation theorem, coupling into each of the mechanical modes $\hat{x}$ and optical modes $\hat{E}$.
        At low frequencies---up to $\sim$kHz---contributions to the mechanical motion of the test masses are plotted as power spectra.
        At these frequencies, thermal noise (red) dominates over the zero-point fluctuations of the test masses (purple).
        At high frequencies---in the $\sim$THz range---the optical power spectrum is shown.
        Here, the optical vacuum fluctuations (purple) are the relevant effect; the thermal occupation of optical modes (red) is exponentially suppressed.
        For noise each curve, the relevant prefactors to the susceptibilities $\Im \chi$ [from \cref{eq:QN}; \cref{eq:TN,eq:boseLimits}] are also shown.
        Gray shows the optical sidebands due to mechanical motion.
    }
\end{figure*}

\emph{Quantum fluctuation-dissipation theorem.}---A system in thermal equilibrium at temperature $T$ can be modeled by the coupling of its observables to a noisy force from the environment.
In the simplest case of a single observable $\hat{x}$, this coupling can be described by an interaction Hamiltonian $\hat{H}_{\t{int}} = -\hat{x}\hat{f}_x$, where $\hat{f}_x$ is the generalized force conjugate to the system operator $\hat{x}$ originating from the system's quantum and thermal environmental fluctuations.
In the linear response regime, the quantum fluctuation-dissipation theorem (FDT) states that the (symmetrized double-sided) power spectral density of any system observable $\hat{y}$ is~\cite[Eq. (7.2)]{Kubo66}
\begin{equation}
    \bar{S}_{yy} [\omega] = \hbar \coth\left( \frac{\hbar \omega}{2 k_B T} \right) \Im \chi_{yx}[\omega],
\end{equation}
where $\chi_{yx}$ is the susceptibility of the observable $\hat{y}$ to the generalized force $\hat{f}_x$, i.e. $\hat{y}[\omega] = \chi_{yx}[\omega] \hat{f}_x[\omega]$.
Using the identity $\coth(\tfrac{\alpha}{2}) = 1 + 2(\ee^\alpha -1)^{-1}$, we rewrite this as
\begin{equation}\label{eq:fdt}
	\bar{S}_{yy}[\omega] = \hbar \left( 2n_\t{th}[\omega] + 1 \right) \Im \chi_{yx}[\omega],
\end{equation}
where $n_\t{th}[\omega] \equiv (\ee^{\hbar \omega/k_B T}-1)^{-1}$ is the Bose-Einstein occupation number.

We define the quantum noise (QN) in $\hat{y}$ to be its fluctuations at zero temperature:
\begin{equation}\label{eq:QN}
    \bar{S}_{yy}^\t{QN}[\Omega] \equiv \lim_{T\rightarrow 0} \bar{S}_{yy}[\omega] = \hbar \Im \chi_{yx}[\omega],
\end{equation}
also called the zero-point fluctuations.

The thermal noise (TN) is then the remaining $T$-dependent term in \cref{eq:fdt},
\begin{equation}\label{eq:TN}
    \bar{S}_{yy}^\t{TN}[\omega] \equiv \hbar \cdot 2n_\t{th}[\omega] \Im \chi_{yx}[\omega].
\end{equation}
Indeed, in the regime where the thermal energy dominates ($k_B T \gg \hbar \omega$), we have $n_\t{th} \approx k_B T/\hbar \omega \gg 1$ and recover the classical FDT result, $\bar{S}_{yy} \approx \bar{S}_{yy}^\t{TN} \approx (2k_B T/\omega) \Im \chi_{yx}$.

The power of the FDT is that mere knowledge of the susceptibility---an object accessible to classical experimenters---dictates all fundamental (i.e. quantum and thermal) noises of interest.
Even further, it implies that the thermal and quantum noises are directly related to each other as
\begin{equation}\label{eq:SyyTNQN}
    \bar{S}_{yy}^\t{TN}[\omega] = 2n_\t{th}[\omega] \bar{S}_{yy}^\t{QN}[\omega].
\end{equation}
Thus, one can be bootstrapped from the other, even without direct knowledge of the susceptibility.

For a system in either the ``cold'' or ``hot'' regime, we can approximate the occupation number as
\begin{equation}\label{eq:boseLimits}
	n_\t{th}[\omega] \approx 
    \begin{dcases}
    	\ee^{-\hbar \omega/k_B T}, & k_B T \ll \hbar \omega \text{ (``cold'')},\\
    	\frac{k_B T}{\hbar \omega}, & k_B T \gg \hbar \omega \text{ (``hot'')},
    \end{dcases}
\end{equation}
in \cref{eq:SyyTNQN} to relate the known quantum noise to the thermal noise and vice versa.
In contemporary GW detectors, the mechanical and optical modes are respectively in the hot and cold regimes.
Thus, the known TN in the mechanical degrees of freedom---calculated independently using the classical 
FDT~\cite{Saulson90,Levin98}---can be used to estimate the mechanical QN:
\begin{equation}\label{eq:Syy_mechQN}
    \bar{S}_{yy}^\t{QN,mech}[\omega] \approx \frac{\hbar \omega}{2 k_B T} \bar{S}_{yy}^\t{TN,mech}[\omega].
\end{equation}
Similarly, the known QN in the optical field---calculated independently, say from input-output relations~\cite{Buonanno01,Kimble01}---can be used to estimate the optical TN:
\begin{equation}\label{eq:Syy_optTN}
    \bar{S}_{yy}^\t{TN,opt}[\omega] \approx 2 \ee^{-\hbar \omega/k_B T} \cdot \bar{S}_{yy}^\t{QN,opt}[\omega].
\end{equation}
\Cref{fig:spectrum} qualitatively shows the well-understood mechanical TN and optical QN, as well as the bootstrapped mechanical quantum and optical thermal noises.
In the following we discuss the specifics of each of the mechanical and optical degrees of freedom in GW detectors.

\emph{Test-mass quantization.}---The test masses in GW detectors are engineered to be acoustic frequency mechanical oscillators.
Their simplest description is through a lumped element model of a mechanical force $\hat{f}_x$---by definition conjugate to the displacement $\hat{x}$---driving the displacement, i.e. $\hat{x}[\Omega] = \chi_{xx}[\Omega] \hat{f}_x[\Omega]$.
Given the test mass pendulum mode is structurally damped, the damping rate is $\Gamma_m[\Omega] = \Omega_m^2 / \Omega Q$, where $\Omega_m$ is the mechanical resonance frequency and $Q$ the mode quality factor~\cite{Saulson90}.
The pendulum mode susceptibility is then
\begin{equation}\label{eq:chixx}
    \chi_{xx}^{-1}[\Omega] = m(-\Omega^2 +\Omega_m^2 - i \Omega_m^2 / Q). 
\end{equation}
It is precisely the interaction of the test mass oscillator with its environment---and the concomitant spreading of its susceptibility in frequency---that was assumed to be negligible in the analysis of Braginsky \etal~\cite{Braginsky03}.
Accounting for it consistently using the quantum FDT shows that the zero-point motion of the oscillator is the mechanical quantum noise:
\begin{equation}\label{SxxQNsingle}
 	\bar{S}_{xx}^\t{QN}[\Omega] = \frac{\hbar}{m}\cdot 
 		\frac{\Omega_m^2/Q}{(\Omega^2 - \Omega_m^2)^2 + (\Omega_m^2 / Q)^2}.
\end{equation} 
Even in this simple model, the quantum noise of the test mass is a broadband displacement noise~\cite{Clerk10,Khalili2012} that cannot, prima facie, be ``filtered'' out as asserted by Braginsky \etal~\cite{Braginsky03}.
In Advanced LIGO for example~\cite{aLIGO2015}, because of the low frequency and low loss of the test mass's pendulum mode ($\Omega_m \approx 2\pi\cdot \SI{0.4}{Hz}$ and $Q\approx 10^{8}$), this model predicts the off-resonant test-mass quantum noise:
\begin{equation*}
	\sqrt{\bar{S}_{xx}^\t{QN}[\Omega \gg \Omega_m]} \approx 10^{-25}\, \t{m/\sqrt{Hz}}\cdot \left(\frac{\Omega}{2\pi\cdot \SI{10}{Hz}}\right)^{-2},
\end{equation*}
six orders of magnitude smaller than the thermal noise.

\begin{figure}[t!]
    \centering
    \includegraphics{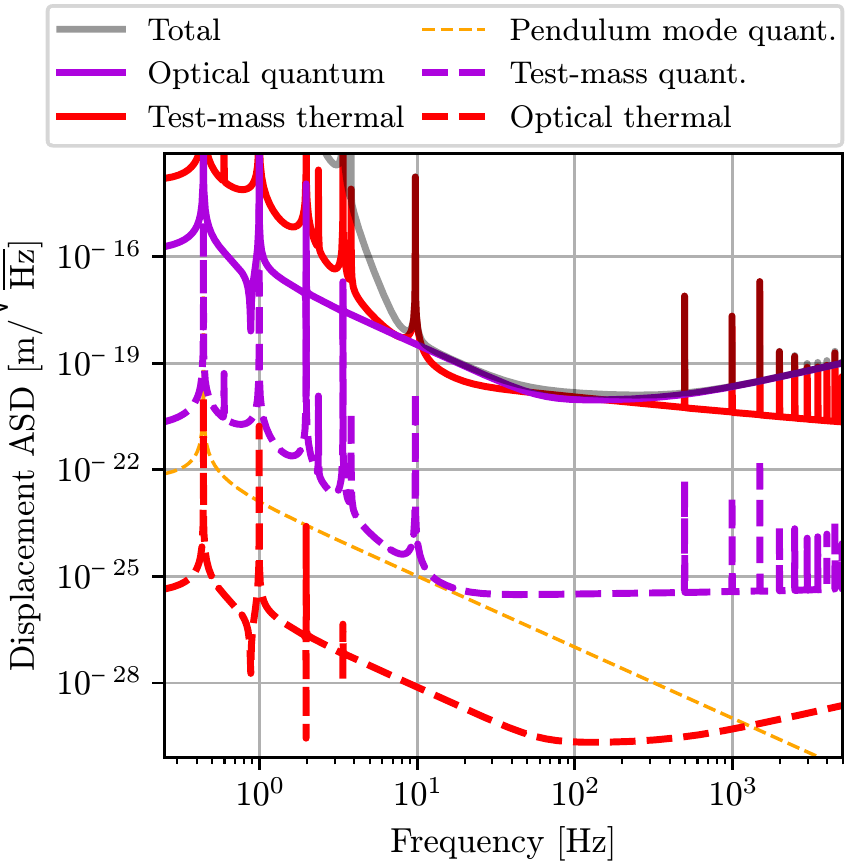}
    \caption{\label{fig:aligo}
    Gray shows the design sensitivity of Advanced LIGO, which is dominated by mechanical thermal noise (red solid) up to $\SI{\sim 200}{Hz}$
    and by optical quantum noise (purple solid) above that frequency~\cite{aLIGO2015}.
    Orange dashed is the predicted mechanical quantum noise in a simplified model of the test-mass pendulum [\cref{SxxQNsingle}],
    while purple dashed shows the prediction [\cref{eq:Syy_mechQN}] for the full mechanical degree of freedom.
    The difference in shape between the test-mass quantum and thermal noises is a result of the frequency-dependent factor in \cref{eq:Syy_mechQN}.
    Red dashed is the optical thermal noise predicted from the known optical quantum noise using \cref{SEEtn}.}
\end{figure}

In reality, the test masses (and their suspensions) are not lumped elements.
They are vibrating elastic continua; further, in interferometric GW detectors, test masses have mirror coatings which have their own elastic fluctuations.
Although a lumped element treatment of the susceptibility is not possible in this case, susceptibilities that describe the thermal noise can nevertheless be derived~\cite{Levin98}.
This is then precisely where our earlier observations are helpful.
Since the relevant frequencies $\Omega \approx 2\pi\cdot (0.1 - 10^3)\, \t{Hz}$ and the operating temperature satisfies $T \gg \hbar \Omega/k_B$, these modes are in the ``hot'' regime.
Thus, knowledge of the thermal noise allows a direct and accurate prediction of the broadband mechanical quantum noise using \cref{eq:Syy_mechQN}.
The dashed purple line in \cref{fig:aligo} shows the broadband mechanical quantum noise in Advanced LIGO bootstrapped from the well-modeled mechanical thermal noise (red solid line).
Note that the broadband mechanical quantum noise is relatively white, in contrast to the $1/\Omega$ falloff of the thermal noise amplitude spectral density, a consequence of the frequency prefactor in \cref{eq:Syy_mechQN}.
The dashed orange line is the prediction from a lumped element model of the pendulum mode alone [\cref{SxxQNsingle}].
Clearly, the displacement quantum noise is broadband, but negligible compared to the corresponding thermal noise---a fact that is contingent on the operating temperature.

\emph{Optical thermal noise.}---Information about the motion of the test masses is imprinted onto electromagnetic fields that propagate through the GW detector.
Typically, the incident field has a carrier at frequency $\omega_o$, while all relevant information is contained in field fluctuations at frequency offsets $\Omega$ around the carrier that are small compared to $\omega_o$ (i.e. $\abs{\Omega} \ll \omega_o$).
Thus we are interested in $\bar{S}_{EE}[\omega_o + \Omega]$, which is given by the quantum FDT
\begin{equation*}
    \bar{S}_{EE}[\omega_o + \Omega] = \hbar \left(2 n_\t{th}[\omega_0 + \Omega] + 1 \right) \Im \chi_{EE}[\omega_0 + \Omega].
\end{equation*}
The optical field in current interferometric GWs are in the ``cold'' regime with respect to the optical carrier and ``hot'' with respect to the offset frequency $\Omega$, i.e. $\hbar \Omega \ll k_B T \ll \hbar \omega_o$.
Thus field fluctuations around the carrier are quantified by
\begin{equation*}
    1 + 2 n_\t{th}[\omega_o + \Omega] \approx 1 + 2 \ee^{-\hbar \omega_o/k_B T}
        \left( 1 + \frac{\hbar \Omega}{k_B T}\cdot  \right),
\end{equation*}
which consists of a dominant quantum noise term~\cite{Meers1991,PaceWalls93,Miao19} with an exponentially small thermal noise contribution.
Thus the thermal noise around the electric field carrier is related to the quantum noise by
\begin{equation}\label{SEEtn}
    \bar{S}_{EE}^\t{TN}[\omega_o + \Omega] \approx 2 e^{-\hbar \omega_o/k_B T}\left( 1 + \frac{\hbar \Omega}{k_B T} \right)
        \bar{S}_{EE}^\t{QN}[\omega_o + \Omega].
\end{equation}
In Advanced LIGO, the quantum noise contribution of the optical field fluctuations is well-characterized (see purple solid line in \cref{fig:aligo}).
Applying \cref{SEEtn} allows a direct extrapolation of the optical thermal noise (red dashed line in \cref{fig:aligo}), which is shown to be negligible, even when compared to the already small mechanical quantum noise.

Some designs for future upgrades to detectors and next-generation installations employ cryogenic technologies.
However, the temperature will only be reduced by an order of magnitude or two~\cite{Adhikari2020,Akutsu2019,ETdesign} and the implications of \cref{fig:aligo} will be unchanged.
A contrasting example can be found in superfluid helium-4 Weber bar antennae, which represent an altogether different technology that breaks into a new operational regime.
Here, the (tune-able) mechanical resonance of the superfluid at \SI{\sim 1}{\kilo\hertz} couples to a microwave cavity resonant at \SI{10.6}{\giga\hertz}~\cite{Singh17,DeLorenzo17}.
The microwave readout circuit spans temperatures from \SI{50}{\milli\kelvin} to room temperature.
For temperatures above \SI{0.5}{\kelvin} within this circuit, the microwave modes enter the hot regime and its thermal noise becomes significant compared to quantum fluctuations.

\emph{Opto-/electro-mechanical interactions.}---\Cref{fig:aligo} shows the noises that contribute to the free-running displacement $\hat{x}_0$ of the test masses, where the effect of optical and electrical feedback has been removed by the calibration process.
Here, we explain why the specifics of such feedback do not affect our analysis as far as metrology is concerned.

The feedback force on the oscillator can be written as
\begin{equation*}
    \hat{f}_\t{fb}[\Omega] = \chi_{xx,\t{fb}}^{-1}[\Omega]\hat{x}[\Omega],
\end{equation*}
where $\chi_{xx,\t{fb}}^{-1}$ is the displacement-to-force open-loop transfer function, and we have omitted the noise component of the feedback since it has no effect on calibration.
This feedback may be electro-optic control loops or direct optical feedback from detuned interaction~\cite{Kom22}.
The displacement fluctuations due to the FDT [\cref{eq:fdt}] can equivalently be written as force fluctuations $\hat{f}_x$, with spectrum
\begin{equation}\label{eq:fdt_force}
    \bar{S}_{f_xf_x}[\Omega] = \hbar \left( 2n_\t{th}[\Omega] + 1 \right) \Im \left(- \chi_{yx}^{-1}[\Omega]\right),
\end{equation}
that sum with the feedback force $\hat{f}_\t{fb}$.

As a result of the feedback, the test-mass susceptibility is modified from its intrinsic form $\chi_{xx}$ [see \cref{eq:chixx}] to an effective (i.e. closed-loop) susceptibility, $\chi_{xx,\t{cl}}~\equiv~\chi_{xx} / (1~-~\chi_{xx} \chi_{xx,\t{fb}}^{-1})$.
The displacement observed with the loop closed is then $\hat{x}_\t{cl}~\equiv~\chi_{xx,\t{cl}} \hat{f}_x$.
This can be extended to also include the GW signal, which couples to the test mass displacement via a force $\hat{f}_\t{gw}$~\cite{Weber1957}.
The free-running displacement is then inferred as
\begin{equation*}
    \hat{x}_0 = \hat{x}_\t{cl} \left(1 - \chi_{xx} \chi_{xx,\t{fb}}^{-1}\right) = \chi_{xx} \left(\hat{f}_x + \hat{f}_\t{gw}\right),
\end{equation*}
where we drop the frequency-dependence of each term for brevity.
By convention, the spectrum $\bar{S}_{ff}$ of the noise $\hat{f}_x$ is calibrated to a displacement spectrum and then plotted as in \cref{fig:aligo}.
Since the effect of the feedback is common to all forces, our ability to measure $\hat{f}_\t{gw}$ depends only on the force noise $\hat{f}_x$ and not on the behavior of the feedback system.
In the normal operation of Advanced LIGO, additional technical noises dominate over these fundamental noise sources at low frequencies ($\lesssim \SI{10}{\hertz}$ for the Advanced LIGO design)~\cite{Buik20}, but we have omitted these for simplicity.

Our conclusion that the effect of feedback is inconsequential for metrology (as in GW detectors) should not be 
confused with a statement on feedback-based quantum state preparation in general.
For example, feedback can be used, given certain conditions on the measurement sensitivity, to trap and cool the 
motion of test masses, as is done to the pendulum mode in Ref.~\cite{Whitt21}. 
For the purposes of metrology, however, such an exercise will suppress the signal 
(i.e. the force originating from GWs $\hat{f}_\t{gw}$) and offer no improvement in signal-to-noise ratio.

\emph{Conclusion.}---Thermal and quantum noises place fundamental limits on sensitivities achievable by GW detectors.
In this letter, we expand on Braginsky's treatment of these noise sources~\cite{Braginsky03} using the general fluctuation-dissipation theorem.
Our approach allows a direct computation of mechanical quantum noise (``test-mass quantization noise'') and optical thermal noise from the well-understood mechanical thermal noise and optical quantum noise respectively.
In doing so we settle the long-standing question of test-mass quantization noise in GW detectors: it is a broadband source of noise that cannot be neglected on the grounds of being limited to certain frequencies, but it lies many orders of magnitude below the sensitivity of any GW detector based \mbox{on current technology}.

\emph{Acknowledgments.}---The authors acknowledge the support of the National Science Foundation and the LIGO Laboratory.
LIGO was constructed by the California Institute of Technology and Massachusetts Institute of Technology with funding from the National Science Foundation, and operates under Cooperative Agreement No. PHY-1764464.
The authors additionally thank Lisa Barsotti, Joe Bentley, Farid Khalili, Mikhail Korobko, Sergey Vyatchanin, Bernard Whiting and Christopher Wipf for useful comments.
This paper has LIGO Document Number LIGO-P2200369.

\bibliography{refs_fdt}
\bibliographystyle{apsrev4-2.bst}

\end{document}